\documentstyle[12pt]{article}
\begin{document}
\newcommand{\be}{\begin{equation}}
\newcommand{\ee}{\end{equation}}
\newcommand{\ba}{\begin{array}}
\newcommand{\ea}{\end{array}}
\newcommand{\bea}{\begin{eqnarray}}
\newcommand{\eea}{\end{eqnarray}}
\begin{center}
{\Large{\bf The Coulomb-Oscillator Relation on
\\[5mm]
n-Dimensional Spheres and Hyperboloids}}

\vspace{0.5cm}
{\bf E.G.Kalnins}

\vspace{0.3cm}
{\it Department of Mathematics and Statistics, University of Waikato,
Hamilton, new Zealand.}

\vspace{0.5cm}
{\bf W.Miller, Jr.}

\vspace{0.3cm}
{\it School of Mathematics, University of Minnesota, Minneapolis,
Minnesota, 55455, U.S.A.}

\vspace{0.5cm}
{\bf G.S.Pogosyan}\footnote{Permanent address:
Laboratory of Theoretical Physics, Joint Institute for Nuclear
Research, Dubna, Russia and International Center for Advanced Studies,
Yerevan State University, Yerevan, Armenia}

\vspace{0.3cm}
{\it Centro de Ciencias F\'{\i}sicas
Universidad Nacional Aut\'onoma de M\'exico
Apartado Postal 48--3 \\ 62251 Cuernavaca, Morelos, M\'exico}

\end{center}

\vspace{0.6cm}

\begin{center}
\noindent{\large {\bf Abstract}}
\end{center}

\vspace{0.3cm}

\noindent
In this paper we establish a relation between Coulomb
and oscillator systems on  $n$-di\-men\-si\-o\-nal spheres and
hyperboloids
for $n\geq 2$. We  show that, as in Euclidean space, the quasiradial
equation for the $n+1$ dimensional Coulomb problem coincides with the
$2n$-dimensional
quasiradial oscillator equation on  spheres and hyperboloids.
Using the solution of the Schr\"odinger equation for the oscillator
system, we construct the energy spectrum and wave functions for the
Coulomb problem.

\section{Introduction}

It has a long been known that the Coulomb and oscillator
potentials are two paradigms in quantum mechanics that possesses
dynamical or hidden symmetries: $O(n+1)$ for motion in a Coulomb
field \cite{BI} and $SU(n)$ for the oscillator. On the other hand the
connections with
these two Lie groups of dynamical symmetries provide  relations
between the Coulomb and oscillator systems.  In particular  the $(n+1)$
radial Schr\"odinger equation for the Coulomb system is identical
to the oscillator equation for $2n$-dimensions by the duality
transformation
\cite{TER-ANT}. It is also known that the complete relation (not
only for the radial part) is possible for only special dimensions (2,2),

(3,4) and (5,8) respectively. The dual mappings in these cases are
so-called Levi-Civita, Kustaanheimo-Stiefel and Hurwitz transformations
\cite{LC,KS,DMPST}.

The generalization of the Coulomb problem to the three-sphere
has been done in the famous article of Schr\"odinger  \cite{SCHR}
and for the $n$-dimensional hyperboloid \cite{IS}. Later the
Coulomb and oscillator problem on  spheres and
pseudospheres was discussed from many point of view in
\cite{HIG,LEEM,BOK,BIJ,GRO1,GRO2,MPSV,DAS,GROP1,GROP2,KMP0,KMP1}

In a previous article \cite{KMP2} we have constructed a series of
complex mappings $S_{2C}\rightarrow S_2$, $S_{4C}\rightarrow S_3$
and $S_{8C}\rightarrow S_5$, which extend to spherical geometry the
Levi-Civita, Kustaanheimo-Stiefel and Hurwitz transformations, well
known for Euclidean space. We have shown that these transformations
establish a correspondence between Coulomb and oscillator
problems in classical and quantum mechanics for dimensions (2,2),
(3,4) and (5,8) on the spheres. A detailed analysis of the real
mapping on the curved space has been done in \cite{NERPOG}. It
was shown that in the stereographic projection (see also the paper
\cite{NERSES}) the relation between Coulomb and oscillator problems
functionally coincide with the flat space Levi-Civita and
Kustaanheimo-Stiefel transformations.

In the present paper we find the relation between the quasiradial
Schr\"odinger equations for Coulomb and oscillator problems
on the $n$-dimensional sphere and two-sheeted hyperboloids for $n\geq
2$.

\section{Coulomb-oscillator relation on n-sphere}

The Schr\"odinger equation describing the nonrelativistic quantum motion

on the $n$-di\-men\-si\-onal sphere: $s_0^2+s_1^2+\cdots + s_n^2=R^2$,
where $s_i$ are Cartesian coordinates in ambient Euclidean
$(n+1)$-space,
has the following form ($\hbar=\mu=1$)
\begin{eqnarray}
\label{SCH1}
{\cal H} \Psi =
\left[ - \frac{1}{2} \Delta_{LB} + V({\vec s})\right]\Psi
= E \Psi
\end{eqnarray}
where the Laplace-Beltrami operator in arbitrary curvilinear coordinates

$\xi_\mu$ is
\begin{eqnarray}
\Delta_{LB} = \frac{1}{\sqrt{g}} \frac{\partial}{\partial \xi_\mu}
g^{\mu\nu} \sqrt{g}
\frac{\partial}{\partial\xi_\nu},
\qquad g=\det ||g_{\mu\nu}||,
\qquad
g_{\alpha\mu}g^{\mu\nu} = \delta^{\nu}_{\alpha}.
\end{eqnarray}
For  any central potential $V({\chi})$ the Schr\"odinger equation
admits  separation of variables in hyperspherical coordinates
\begin{eqnarray*}
s_0 &=& R\cos\chi \\
s_1 &=& R\sin\chi\cos\vartheta_1 \\
s_2 &=& R\sin\chi\sin\vartheta_1\cos\vartheta_2 \\
&...& \\
s_{n-1} &=& R\sin\chi\sin\vartheta_1 \sin\vartheta_2 \cdots
\sin\vartheta_{n-2}\cos\varphi \\
s_{n}   &=& R\sin\chi\sin\vartheta_1\sin\vartheta_2 \cdots
\sin\vartheta_{n-2}\sin\varphi.
\end{eqnarray*}
where $\chi, \vartheta_1, ... \vartheta_{n-2} \in [0,\pi]$,
$\varphi\in [0, 2\pi)$. We can separate the angular part of the wave
function using the ansatz
\begin{eqnarray}
\Psi(\chi, \vartheta_1, \cdots \vartheta_{n-2}, \varphi) =
{\cal R}(\chi) \,
Y_{Ll_1,l_2,l_{n-2}}(\vartheta_1, \cdots \vartheta_{n-2}, \varphi)
\end{eqnarray}
where $l_i$ are the angular hypermomenta and $L$ is total angular
momentum,
and the hyperspherical function
$Y_{L,l_1,l_2,l_{n-2}}(\vartheta_1, \cdots \vartheta_{n-2}, \varphi)$
is the solution of the Laplace-Beltrami eigenvalue equation on the
$n-1$ dimensional sphere. After separation of variables in
(\ref{SCH1}) we obtain the quasiradial equation
\begin{eqnarray}
\frac{1}{\sin^{n-1}\chi}
\frac{d}{d\chi} \sin^{n-1}{\chi}
\frac{d {\cal R}(\chi)}{d\chi} +
\left[2R^2E - \frac{L(L+n-2)}{\sin^2{\chi}}
- 2R^2 V(\chi) \right]\,{\cal R}(\chi) = 0.
\end{eqnarray}
Using  the substitution
\begin{eqnarray}
\label{OSC1}
Z(\chi)  = (\sin\chi)^{\frac{n-1}{2}}{\cal R}(\chi)
\end{eqnarray}
we find
\begin{eqnarray}
\label{SCH2}
\frac{d^2 Z}{d\chi^2} + \left[{\tilde E} -
\frac{(2L+n-1)(2L+n-3)}{4\sin^2{\chi}} - 2R^2 V(\chi) \right]
\, Z = 0
\end{eqnarray}
where ${\tilde E} = 2R^2E + (n-1)^2/{4}$ and the quasiradial
wave function $Z(\chi)$ satisfies the normalization condition
\begin{eqnarray}
\int_{0}^{\pi} Z(\chi) Z^{*}(\chi)\, R^n\,  d\chi = 1.
\end{eqnarray}

\noindent
{\bf 2.1.}
Let us now consider the n-dimensional oscillator potential
\cite{HIG,LEEM}
\begin{eqnarray}
V(\chi) =  \frac{\omega^2 R^2}{2}
\frac{s_1^2 + s_2^2 +..... +s_{n}^2}{s_0^2}
= \frac{\omega^2 R^2}{2} \tan^2\chi.
\end{eqnarray}
Substituting the oscillator potential in equation (\ref{SCH2})
we obtain the P\"oschl-Teller type equation
\begin{eqnarray}
\label{OSCIL1}
\frac{d^2 Z}{d\chi^2} + \left[\epsilon -
\frac{\nu^2-\frac14}{\cos^2\chi} -
\frac{(L+\frac{n-2}{2})^2-\frac14}{\sin^2{\chi}}\right]\, Z = 0
\end{eqnarray}
where $\nu = \sqrt{\omega^2 R^4+1/4}$ and $\epsilon = {\tilde E}
+ \omega^2 R^4$.
The solution of the above equation, regular for
$\chi\in [0,\pi/2]$ and expressed in terms of the hypergeometric
function, is \cite{FLUG}
\begin{eqnarray}
\label{OSCFUN1}
Z(\chi) \equiv Z_{n_r L \nu}^{n}(\chi) =
\sqrt{\frac{2(2n_r+L+\nu+\frac{n}{2})
\Gamma(n_r+L+\nu+\frac{n}{2})\Gamma(n_r+L+\frac{n}{2})}
{R^{n}\, [\Gamma(L+\frac{n}{2})]^2 \Gamma(n_r+\nu+1)
(n_r)!}}
\nonumber\\[3mm]
(\sin\chi)^{L+\frac{n-1}{2}}
\,
(\cos\chi)^{\nu+\frac12}
\,
{_2F_1}(-n_r, n_r+L+\nu+\frac{n}{2};
\, L+\frac{n}{2}; \, \sin^2\chi),
\end{eqnarray}
and the $\epsilon$ is quantized as
\begin{eqnarray}
\label{ENER1}
\epsilon = (2n_r+ L + \nu + \frac{n}{2})^2
\end{eqnarray}
where $n_r+L=0,1,2,...$ is a ``quasiradial" quantum number.
The energy spectrum of the $n$-dimensional oscillator is given by
\begin{eqnarray}
\label{ENER11}
E_{N}^{n}(R) = \frac{1}{2R^2}\left[(N+1)(N+n) +
(2\nu-1)(N + \frac{n}{2})\right]
\end{eqnarray}
where $N=2n_r +L = 0,1,...$ is principal quantum number.
In the contraction limit $R\to\infty$, $\chi\to0$ and
$R\chi \sim r - $fixed and $\nu\sim \omega R^2$, we see that
\begin{eqnarray}
\label{LIM-ENER1}
\lim_{R\to\infty}E_{N}^{n}(R) = \omega (N + \frac{n}{2})
\end{eqnarray}
and
\begin{eqnarray}
\label{LIMIT-OSCFUN1}
\lim_{R\to\infty}\,
(R)^{\frac{n-1}{2}}
\,
Z_{N L \nu}^{n}(\chi) =
\frac{(\omega)^{\frac{L}{2}+\frac{n}{4}}}
{\Gamma(L+\frac{n}{2})}
\,
\sqrt{\frac{2\Gamma(\frac{N+L+n}{2})}
{(\frac{N-L}{2})!}}
\
r^{L+\frac{n-1}{2}}
\,
e^{-\frac{\omega r^2}{2}}
\,
{_1F_1}(-\frac{N-L}{2}, \, L+\frac{n}{2}; \, \omega r^2).
\end{eqnarray}
Formula (\ref{LIMIT-OSCFUN1}) coincides with the known formula for
 n-dimensional flat radial wave functions \cite{PST}.

\noindent
{\bf 2.2.}
The potential, which is the analogue of the the Coulomb potential
on the n-dimen\-sio\-nal sphere, has the following form
\cite{SCHR,HIG,LEEM}:
\begin{eqnarray}
V(\chi) = - \frac{\alpha}{R}
\frac{s_0}{\sqrt{s_1^2 + s_2^2 +..... + s_{n}^2}}
= - \frac{\alpha}{R} \cot\chi.
\end{eqnarray}
The Schr\"odinger equation (\ref{SCH2}) for this potential is
\begin{eqnarray}
\label{KC1}
\frac{d^2 Z}{d\chi^2} + \left[{\tilde E} -
\frac{(2L+n-1)(2L+n-3)}{4\sin^2{\chi}} + 2\alpha R \cot\chi
\right]\, Z = 0.
\end{eqnarray}
We make now a transformation to the new variable
$\theta \in [0, \frac{\pi}{2}]$
\begin{eqnarray}
\label{TRANS1}
e^{i\chi} = \cos\theta,
\end{eqnarray}
which is possible if we continue the variable $\chi$ in the complex
domain {\bf G}: Re\ $ \chi = 0$, $0\leq$\ Im\ $\chi < \infty$
(see Fig.1). We complexify also the coupling constant $\alpha$
by putting $k=i\alpha$ such that
\begin{eqnarray}
\label{TRANS2}
\alpha \cot\chi =  k (1- 2 \sin^{-2}\theta).
\end{eqnarray}

\vspace{7mm}
\unitlength=1.00mm
\special{em:linewidth 0.4pt}
\linethickness{0.4pt}
\begin{picture}(102.00,126.60)
\put(14.00,44.00){\makebox(0,0)[cc]{0}}
\put(12.00,120.00){\makebox(0,0)[cc]{Im $\chi$}}
\put(102.00,42.00){\makebox(0,0)[cc]{Re $\chi$}}
\put(98.00,50.00){\line(-1,0){91.00}}
\put(65.00,50.00){\line(0,1){67.00}}
\put(65.00,43.00){\makebox(0,0)[cc]{$\pi$}}
\put(65.00,119.00){\line(-1,0){4.00}}
\put(59.00,119.00){\line(-1,0){4.00}}
\put(52.00,119.00){\line(-1,0){5.00}}
\put(44.00,119.00){\line(-1,0){5.00}}
\put(36.00,119.00){\line(-1,0){5.00}}
\put(28.00,119.00){\line(-1,0){5.00}}
\put(41.00,86.00){\makebox(0,0)[cc]{{\bf G}}}
\put(74.00,86.00){\makebox(0,0)[cc]{G}}
\put(96.00,50.00){\vector(1,0){2.00}}
\put(39.00,50.00){\vector(1,0){3.00}}
\put(65.00,84.00){\vector(0,1){1.00}}
\put(44.00,119.00){\vector(-1,0){2.00}}
\put(20.00,86.00){\vector(0,-1){2.00}}
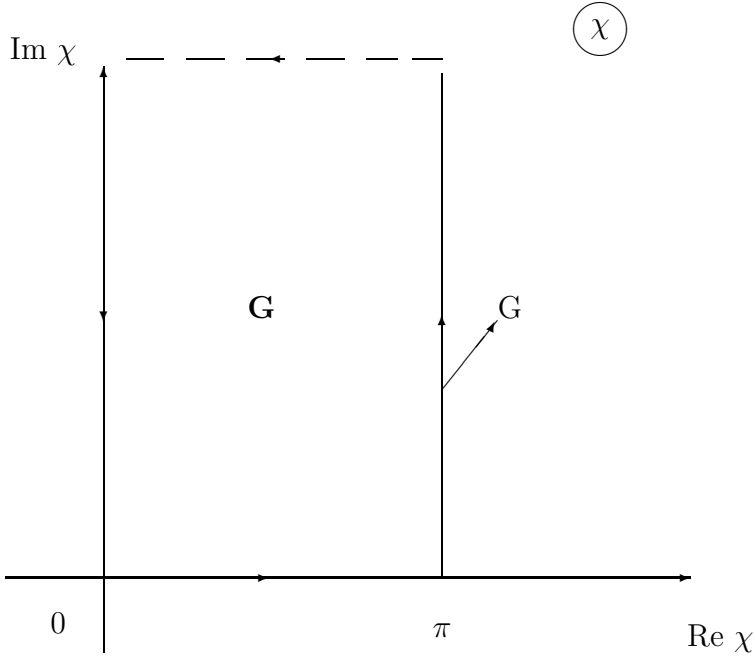
\put(-5.00,22.00){\makebox(0,0)[lc]{{\bf Figure 1:}\ \  Domain {\bf G}
=\{ $0\leq$ Re $\chi\leq \pi$; $0\leq$ Im $\chi< \infty$\} on the
complex plane of $\chi$.}}
\put(20.00,40.00){\line(0,1){78.00}}
\put(86.00,123.00){\circle{7.21}}
\put(86.00,123.00){\makebox(0,0)[cc]{$\chi$}}
\put(65.00,75.00){\line(4,5){7.33}}
\put(71.00,82.00){\vector(1,2){1.00}}
\put(20.00,118.00){\vector(0,1){0.00}}
\end{picture}

\vspace{-7mm}
\noindent
As result we obtain the equation
\begin{eqnarray}
\label{COOS1}
\frac{d^2 W}{d\theta^2} + \left[\epsilon -
\frac{\nu^2 - \frac{1}{4}}{\cos^2\theta} -
\frac{(2L+n-2)^2-\frac14}{\sin^2\theta} \right]\, W = 0
\end{eqnarray}
where $W(\theta) = ({\cot\theta})^{\frac12} Z(\theta)$ and
\begin{eqnarray}
\label{ENER2}
\epsilon = {\tilde E} + 2kR,
\qquad
\nu^2 =  {\tilde E} - 2kR.
\end{eqnarray}
>From above equation we see that, up to the substitution (\ref{ENER2})
and transformation $L \rightarrow 2L$, the quasiradial equation
(\ref{COOS1}) for the $n^{coul}=(d+1)$-dimensional Coulomb problem
coincides with the $n^{osc}=2d$-dimensional quasiradial oscillator
equation (\ref{OSCIL1}). This means that relations between these two
systems are possible only for oscillators in even dimensions:
$n^{osc}=2,4,6,8...$.

Thus equation (\ref{COOS1}) describes the $2(n-1)$ - dimensional
oscillator quasiradial functions with  even angular momentum $2L$.
The regular, for $\theta\in [0,\pi/2]$ and $\nu \leq 1/4$, solution of
this
equation according to (\ref{OSCFUN1}) has the form
\begin{eqnarray}
\label{OSCFUN2}
Z(\theta)
=
\frac{W(\theta)}{\sqrt{\cot\theta}} \equiv
Z_{n_r L}(\theta)
&=&
C_{n_r L}^n(\nu)
\,
(\sin\theta)^{2L+n-1}
\,
(\cos\theta)^{\nu}
\nonumber
\\[2mm]
&\times&
{_2F_1}(-n_r, n_r+2L+\nu+n-1; \, 2L+n-1; \, \sin^2\theta)
\end{eqnarray}
where $C_{n_r L}^n(\nu)$ is the normalization constant. To compute the
constant $C_{n_r L}^n(\nu)$ for the corresponding Coulomb
quasiradial function we require that the wave function (\ref{OSCFUN2})
satisfy the normalized condition
\begin{eqnarray}
R^n\,
\int_{0}^{\pi}
Z_{n_r L} \, Z_{n_r L}^{\diamond} \, d\chi = 1,
\end{eqnarray}
where the symbol ``${\diamond}$" means the complex conjugate together
with the inversion $\chi\rightarrow -\chi$, i.e. $Z^{\diamond}(\chi)
= Z^{*}(-\chi)$. [We choose the scalar product as $Z^{\diamond}$
because for $\chi\in${\bf G} and real $\alpha$, and ${\tilde E}$
the function $Z^{\diamond}(\chi)$ also belongs to the solution space
of (\ref{KC1}).] By  analogy to the work \cite{KMP2} we consider
the integral over contour $G$ in the complex plane of variable
$\chi$ (see Fig.1)
\bea
\label{NOR2}
\oint Z_{n_r L}(\chi) \, Z_{n_r L}^{\diamond}(\chi) d\chi
&=&
\int_{0}^{\pi} Z_{n_r L}(\chi) \, Z_{n_r L}^{\diamond}(\chi) d\chi
+
\int_{\pi}^{\pi+i\infty} Z_{n_r L}(\chi) \,
Z_{n_r L}^{\diamond}(\chi) d\chi
\nonumber\\[2mm]
&+&
\int_{\pi+i\infty}^{i\infty} Z_{n_r L}(\chi) \,
Z_{n_r L}^{\diamond}(\chi) d\chi
+
\int_{i\infty}^0 Z_{n_r L}(\chi) \,
Z_{n_r L}^{\diamond}(\chi) d\chi.
\eea
Using the facts that the integrand  vanishes as $e^{2i\nu\chi}$ and
that  $Z_{n_r L}(\chi)$ is regular in the domain {\bf G} (see Fig.1),
then according to the Cauchy theorem we have
\bea
\label{NOR3}
\int_{0}^{\pi} Z_{n_r L}(\chi) \, Z_{n_r L}^{\diamond}(\chi) d\chi
=
\left(1-e^{2i\pi\nu}\right)\,
\int_0^{i\infty} Z_{n_r L}(\chi) \,
Z_{n_r L}^{\diamond}(\chi) d\chi.
\eea
Making the substitution (\ref{TRANS1}) in the right integral
of eq. (\ref{NOR3}), we find
\bea
\int_{0}^{\pi} Z_{n_r L}(\chi) \,
Z_{n_r L}^{\diamond}(\chi) d\chi
= i\left(1-e^{2i\pi\nu}\right)\,
\int_{0}^{\frac{\pi}{2}}
\left[Z_{n_r L}\right] \tan\theta \, d\theta.
\eea
and after integration over the angle $\theta$ we finally get \cite{BER}
\bea
\label{SSS1}
C_{n_r m}^n(\nu)
=
\sqrt{\frac{(-2i\nu)(\nu+2n_r+2L+n-1) \,(n_r)! \Gamma(2L+n_r+\nu+n-1)}
{R^n [1-e^{2i\pi\nu}](2n_r+2L+n-1)\,(n_r+2L+n-2)!\Gamma(n_r+\nu+1)}}.
\eea
Comparing now the eqs. (\ref{ENER1}) with (\ref{ENER2}) and putting
$k=i\alpha$, we get
\begin{eqnarray}
\label{OMEGA1}
\nu = - \left(n_r+ L + \frac{n-1}{2}\right) + i\sigma,
\qquad
\sigma= \frac{\alpha R}{n_r+ L +\frac{n-1}{2}}
\end{eqnarray}
and obtain the energy spectrum for the Coulomb problem
\begin{eqnarray}
\label{ENERGY1}
E_n = \frac{N(N + n-1)}{2R^2} - \frac{\alpha^2}{2(N+\frac{n-1}{2})^2},
\qquad
N=n_r+L= 0,1,2,....
\end{eqnarray}
Returning  to the variable $\chi$, we see that the Coulomb
quasiradial wave function has the form
\begin{eqnarray}
Z_{NL}(\chi)
&=&
C_{NL}(\sigma) \, (\sin\chi)^{L+\frac{n-1}{2}}
\, \exp[-i\chi(N-L-i\sigma)]
\nonumber\\[3mm]
&\times&
{_2F_1}(-N+L, L+\frac{n-1}{2}+i\sigma; 2L+n-1; 1- e^{2i\chi}),
\end{eqnarray}
where the normalization constant $C_{NL}(\sigma)$ is
\begin{eqnarray}
C_{N L}^n(\sigma) = 2^{L+\frac{n-1}{2}}\, e^{\frac{\pi\sigma}{2}}
\,
\frac{\mid\Gamma(L+\frac{n-1}{2}-i\sigma)\mid}
{\Gamma(2L + n-1)} \,
\sqrt{\frac{[(N+\frac{n-1}{2})^2+\sigma^2](N+L+ n-2)!}
{2R^{n} \pi (N+\frac{n-1}{2}) (N-L)!}}.
\end{eqnarray}

Thus by using the relation between Coulomb and oscillator systems
we have constructed the quasiradial wave functions and energy spectrum
for a  Coulomb system on the n-dimensional sphere.

Finally, note that in the contraction limit $R\rightarrow\infty$
(see for details  \cite{MPSV}) it is easy to recover the well known
formulas for the flat space $n$-dimensional Coulomb problem
both for discrete and continuous spectrum \cite{BI}.

\section{Coulomb-oscillator relation on the n-dimensional two-sheeted
hyperboloid}

The pseudospherical coordinates on n-dimensional two-sheeted
hyperboloid:
$s_0^2-s_1^2-s_2^2-\cdots-s_n^2=R^2$, $s_0\geq R$, are
\begin{eqnarray*}
s_0 &=& R\cosh\tau \\
s_1 &=& R\sinh\tau\cos\vartheta_1 \\
s_2 &=& R\sinh\tau\sin\vartheta_1\cos\vartheta_2 \\
&...& \\
s_{n-1} &=& R\sinh\tau\sin\vartheta_1 \sin\vartheta_2 \cdots
\sin\vartheta_{n-2}\cos\varphi \\
s_{n}   &=& R\sinh\tau\sin\vartheta_1\sin\vartheta_2 \cdots
\sin\vartheta_{n-2}\sin\varphi
\end{eqnarray*}
where $\tau \in [0, \infty)$.
Variables in the Schr\"odinger equation (1) may be separated for any
central potential $V(\tau)$ by the ansatz
\begin{eqnarray}
\Psi(\tau, \vartheta_1, \cdots \vartheta_{n-2}, \varphi) =
{\cal R(\tau)} \,
Y_{Ll_1,l_2,l_{n-2}}(\vartheta_1, \cdots \vartheta_{n-2}, \varphi)
\end{eqnarray}
where, as in previous case $l_i$ are the angular hypermomenta and
$L$ is total angular momentum, and the hyperspherical function
$Y_{L,l_1,l_2,l_{n-2}}(\vartheta_1, \cdots \vartheta_{n-2}, \varphi)$
is the solution of Laplace-Beltrami equation on the $n-1$ dimensional
sphere. After the separation of variables we find the quasiradial
equation
\begin{eqnarray}
\label{SCH22}
\frac{1}{\sinh^{n-1}\tau}
\frac{d}{d\tau} \sinh^{n-1}{\tau}
\frac{d{\cal R}}{d\tau} +
\left[2R^2E - \frac{L(L+n-2)}{\sinh^2{\tau}}
- 2R^2 V(\tau) \right]\, {\cal R} = 0.
\end{eqnarray}
Using now the substitution
\begin{eqnarray}
Z(\tau)  = (\sinh\tau)^{\frac{n-1}{2}} {\cal R}(\tau)
\end{eqnarray}
we come to the equation
\begin{eqnarray}
\label{SCH32}
\frac{d^2 Z}{d\tau^2} + \left[{\tilde E} -
\frac{(2L+n-1)(2L+n-3)}{4\sinh^2{\tau}} - 2R^2 V(\tau) \right]
\, Z = 0
\end{eqnarray}
where ${\tilde E} = 2R^2E - \frac{(n-1)^2}{4}$ and the quasiradial
wave function $Z(\tau)$ satisfy the normalization condition
\begin{eqnarray}
\int_{0}^{\infty} Z(\tau) Z^{*}(\tau)\, R^n\,  d\tau = 1
\end{eqnarray}

\noindent
{\bf 3.1.} The oscillator potential on the two-sheeted n-dimensional
hyperboloid is given by the potential
\begin{eqnarray}
V(\tau) =  \frac{\omega^2 R^2}{2}
\frac{s_1^2 + s_2^2 +..... s_{n}^2}{s_0^2}
= \frac{\omega^2 R^2}{2} \tanh^2\tau.
\end{eqnarray}
>From equation (\ref{SCH32}) we obtain
\begin{eqnarray}
\label{OSCIL11}
\frac{d^2 Z}{d\tau^2} + \left[\epsilon +
\frac{\nu^2-\frac14}{\cosh^2\tau} -
\frac{(L+\frac{n-2}{2})^2-\frac14}{\sinh^2{\tau}}\right]\, Z = 0
\end{eqnarray}
where $\nu = \sqrt{\omega^2 R^4+\frac14}$ and $\epsilon = {\tilde E}
- \omega^2 R^4$.
Thus the oscillator problem on the hyperboloid is described by the
{\it modified  P\"oschl-Teller} equation and, unlike  the oscillator
equation on the sphere which has only bound spectrum, the equation
(\ref{OSCIL11}) possesses both bound and unbound states.

The discrete wave-functions regular on the line $\tau \in [0, \infty)$,
have the form \cite{GROP1,KMP1,LANGE}
\begin{eqnarray}
Z(\tau)
&\equiv&
Z_{n_r L}(\tau) =
\frac{1}{\Gamma(L+\frac{n}{2})}
\,
\sqrt{\frac{2(\nu-L-2n_r-\frac{n}{2})\Gamma(\nu-n_r)
\Gamma(n_r+L+\frac{n}{2})}
{R^n (n_r)!\Gamma(\nu-L-n_r-\frac{n}{2}+1)}}
\nonumber\\[3mm]
&\times&
(\sinh\tau)^{L+\frac{n-1}{2}}
\,
(\cosh\tau)^{2n_r-\nu+\frac12}
\,
{_2F_1}(-n_r, -n_r+\nu; \, L+\frac{n}{2}; \, \tanh^2\tau),
\end{eqnarray}
with $n_r=0,1,...[\frac12(\nu-L-\frac{n}{2})]$. The $\epsilon$
is quantized by
\begin{eqnarray}
\label{EPSILON21}
\epsilon = - (2n_r + L- \nu + n/2)^2
\end{eqnarray}
and the energy spectrum for the quantum oscillator on the
$n$-dimensional two-sheeted hyperboloid takes the value
\begin{eqnarray}
\label{ENER21}
E_{N}^{n}(R) = \frac{1}{2R^2}\left[-N(N+n-1) +
(2\nu-1)(N + \frac{n}{2})\right].
\end{eqnarray}
Here $N=2n_r+L$ is a principal quantum number and the bound state
solution is possible only for
\begin{eqnarray}
\label{BOUND1}
0\leq N \leq  \left[\nu-\frac{n}{2}\right]
\end{eqnarray}
In the contraction limit $R\to\infty$, $\tau\sim r/R$ and
$\nu\sim \omega R^2$ we see that the continuous spectrum is vanishing
while the discrete spectrum is infinite, and it is easy to reproduce
the oscillator energy spectrum (\ref{LIM-ENER1}) and wave
function (\ref{LIMIT-OSCFUN1}).

\vspace{0.3cm}
\noindent
{\bf 3.2.}
The Coulomb potential on the two-sheeted n-dimensional
hyperboloid has the form \cite{IS,GRO1}
\begin{eqnarray}
\label{COULOMB1}
V(\tau) = - \frac{\alpha}{R}
\left(\frac{s_0}{\sqrt{s_1^2 + s_2^2 +.....+ s_{N}^2}} - 1\right)
= - \frac{\alpha}{R}(\coth\tau -1).
\end{eqnarray}
Substituting  potential (\ref{COULOMB1}) in Schr\"odinger equation
(\ref{SCH32}) we arrive at
\begin{eqnarray}
\label{COOS21}
\frac{d^2 Z}{d\tau^2} + \left[({\tilde E}- 2\alpha R) -
\frac{(2L+n-1)(2L+n-3)}{4\sinh^2{\tau}} + 2\alpha R \coth\tau
\right]\, Z = 0,
\end{eqnarray}
which is  known as the {\it Manning-Rosen} potential problem \cite{MR}.

Making the transformation from variable  $\tau$
($0\leq \tau < \infty$), to the new variable $\mu \in [0, \infty)$
\begin{eqnarray}
\label{TRANS21}
e^{\tau} = \cosh\mu,
\end{eqnarray}
and setting  $Z(\mu) = W(\mu)/\sqrt{\coth\mu}$,
we go to the modified P\"oschl-Teller equation
\begin{eqnarray}
\label{COOS22}
\frac{d^2 W}{d\mu^2} + \left[{\tilde E} +
\frac{(-{\tilde E}+ 4\alpha R) - \frac{1}{4}}{\cosh^2\mu} -
\frac{(2L+n-2)^2-\frac14}{\sinh^2\mu} \right]\, W = 0.
\end{eqnarray}
It can be see  from the  eq.(\ref{COOS22}) with
the substution
\begin{eqnarray}
\label{OSC-COUL2}
\epsilon = {\tilde E}, \qquad \nu^2 = - {\tilde E} + 4\alpha R.
\end{eqnarray}
and the transformation $L \rightarrow 2L$, the quasiradial equation
(\ref{COOS1}) for $n^{coul}=2d+1$-dimensional Coulomb problem coincides
with the $n^{osc}=2d$ - dimensional quasiradial oscillator equation
(\ref{OSCIL11}).

Thus the regular for $\mu\in [0,\infty)$ solution of equation
(\ref{COOS21})
or (\ref{COOS22}) has the form
\begin{eqnarray}
\label{OSCFUN22}
Z(\mu)
=
\frac{W(\mu)}{\sqrt{\coth\mu}} \equiv
Z_{n_r L}^n(\mu)
&=&
A_{n_r L}^n(\nu)
\,
(\sinh\mu)^{L+\frac{n}{2}}
\,
(\cosh\mu)^{2n_r-\nu}
\nonumber
\\[2mm]
&\times&
{_2F_1}\left(-n_r, -n_r+\nu; \, L+\frac{n}{2}; \, \tanh^2\mu\right)
\end{eqnarray}
where $A_{n_r L}^n(\nu)$ is the normalization constant. The constant
$A_{n_r L}^n(\nu)$ is computed from the requirement that the wave
function (\ref{OSCFUN22}) satisfies the normalized condition
\begin{eqnarray}
R^n\,
\int_{0}^{\infty}
|Z_{n_r L}^n(\tau)|^2 \, d\tau =
R^n\,
\int_{0}^{\infty}
|Z_{n_r L}^n(\mu)|^2 \, \tanh\mu \, d\mu = 1.
\end{eqnarray}
and has the following form
\begin{eqnarray}
\label{COUL-CONST21}
A_{n_r L}^n(\nu) =
\frac{1}{\Gamma(L+\frac{n}{2})}
\,
\sqrt{\frac{2\nu(\nu-L-2n_r-\frac{n}{2})\Gamma(\nu-n_r)
\Gamma(n_r+L+\frac{n}{2})}
{R^n (L+2n_r+\frac{n}{2})(n_r)!\Gamma(\nu-L-n_r-\frac{n}{2}+1)}}.
\end{eqnarray}
Comparing now  eq.(\ref{OSC-COUL2}) with (\ref{EPSILON21})
and passing from the oscillator to the Coulomb angular quantum
number $L\to 2L$ and dimension $n \to 2(n-1)$, we get
\begin{eqnarray}
\nu = (n_r+L+\sigma+ \frac{n-1}{2}),
\qquad
\sigma = \frac{\alpha R}{n_r+L+\frac{n-1}{2}}.
\end{eqnarray}
Thus the discrete energy spectrum of the Coulomb problem on the
$n$-dimensional two-sheeted hyperboloid is described by the formula
\begin{eqnarray}
E_N^n(R) = - \frac{N(N + n-1)}{2R^2} -
\frac{\alpha^2}{2(N+\frac{n-1}{2})^2} + \frac{\alpha}{R},
\end{eqnarray}
where $N=n_r+L$ is the principal quantum number and the bound states
occur for
\begin{eqnarray}
\label{BOUND2}
0\leq N \leq  \left[\sigma-\frac{n-1}{2}\right].
\end{eqnarray}
The discrete wave function has the form
\begin{eqnarray}
\label{COUL-HYPER1}
Z_{N L}^n(\tau)
&=& A_{NL}^n (\sigma) \,
(\sinh\tau)^{L+\frac{n-1}{2}} \, e^{\tau(N-L-\sigma)}
\nonumber
\\[2mm]
&\times&
{_2F_1}\left(-N+L, L+\frac{n-1}{2}+\sigma; \,
2L+n-1; \, 1-e^{-2\tau}\right),
\end{eqnarray}
where the normalization constant $A_{NL}^n(\sigma)$ is
\begin{eqnarray}
\label{NORM-CONST2}
A_{NL}^n (\sigma)
=
\frac{2^{L+\frac{n-1}{2}}}{\Gamma(2L+n-1)}
\,
\sqrt{\frac{[\sigma^2- (N+\frac{n-1}{2})^2]\Gamma(N+L+n-1)
\Gamma(\sigma+L+\frac{n-1}{2})}
{R^n (N+\frac{n-1}{2})(N-L)!\Gamma(\sigma-L-\frac{n-1}{2}+1)}}.
\end{eqnarray}
The solution for the Coulomb quasiradial equation, both
for energy spectrum and wave functions, is identical to that given in
the paper \cite{GRO1} by a path integral approach. We not consider here
the contraction limit $R\to\infty$ to flat $E_n$ Euclidean space for
the Coloumb problem because it has been done already in the same
article \cite{GRO1}.

It should be noted that instead of  substitution (\ref{TRANS21})
it is possible to use the trigonometric transformation
\begin{eqnarray}
e^{-\tau} = \cos\varphi, \qquad\qquad
\varphi \in [0, \pi/2].
\end{eqnarray}
It is easy to see that in this case, up to the permutation
\begin{eqnarray}
\epsilon = - {\tilde E} + 4\alpha R,
\qquad
\nu^2 = - {\tilde E},
\end{eqnarray}
and transformation $L\to 2L$, the quasiradial equation (\ref{COOS21})
for the $n^{coul}=(d+1)$ - dimensional Coulomb problem passes to the
$n^{osc}=2d$ - dimensional quasiradial oscillator equation
(\ref{OSCIL1}).
Thus the Coulomb problem on the two-sheeted hyperboloid is
related to the oscillator problem on the sphere or two-sheeted
hyperboloid.

\section{Coulomb-oscillator relation on the n-dimensional one-sheeted
hyperboloid}

Pseudospherical coordinates on the n-dimensional one-sheeted
hyperboloid:
$s_0^2-s_1^2-s_2^2-\cdots-s_n^2=- R^2$ are
\begin{eqnarray*}
s_0 &=& R\sinh\tau \\
s_1 &=& R\cosh\tau\cos\vartheta_1 \\
s_2 &=& R\cosh\tau\sin\vartheta_1\cos\vartheta_2 \\
&...& \\
s_{n-1} &=& R\cosh\tau\sin\vartheta_1 \sin\vartheta_2 \cdots
\sin\vartheta_{n-2}\cos\varphi \\
s_{n}   &=& R\cosh\tau\sin\vartheta_1\sin\vartheta_2 \cdots
\sin\vartheta_{n-2}\sin\varphi
\end{eqnarray*}
where $\tau \in (-\infty, \infty)$. Variables in Schr\"odinger
equation (1) may be separated using the ansatz
\begin{eqnarray}
\Psi(\tau, \vartheta_1, \cdots \vartheta_{n-2}, \varphi) =
{\cal R(\tau)} \,
Y_{Ll_1,l_2,l_{n-2}}(\vartheta_1, \cdots \vartheta_{n-2}, \varphi)
\end{eqnarray}
where as in the previous case the $l_i$ are the angular hypermomenta,
$L$ is total angular momentum, and the hyperspherical function
$Y_{L,l_1,l_2,l_{n-2}}(\vartheta_1, \cdots \vartheta_{n-2}, \varphi)$
is the solution of the Laplace-Beltrami equation on the $n-1$
dimensional
sphere. After  separation of variables we find the quasiradial
equation
\begin{eqnarray}
\label{SCH33}
\frac{1}{\cosh^{n-1}\tau}
\frac{d}{d\tau} \cosh^{n-1}{\tau}
\frac{d{\cal R}}{d\tau} +
\left[2R^2E + \frac{L(L+n-2)}{\cosh^2{\tau}}
- 2R^2 V(\tau) \right]\, {\cal R} = 0.
\end{eqnarray}
Using now the substitution
\begin{eqnarray}
Z(\tau)  = (\cosh\tau)^{\frac{n-1}{2}} {\cal R}(\tau)
\end{eqnarray}
we come to the equation
\begin{eqnarray}
\label{SCHROD42}
\frac{d^2 Z}{d\tau^2} + \left[{\tilde E} +
\frac{(2L+n-1)(2L+n-3)}{4\cosh^2{\tau}} - 2R^2 V(\tau) \right]
\, Z = 0
\end{eqnarray}
where ${\tilde E} = 2R^2E - \frac{(n-1)^2}{4}$ and the quasiradial
wave function $Z(\tau)$ satisfies the normalization condition
\begin{eqnarray}
\int_{-\infty}^{\infty} Z(\tau) Z^{*}(\tau)\, R^n\,  d\tau = 1.
\end{eqnarray}

\noindent
{\bf 4.1}
The oscillator potential on the n-dimensional one-sheeted
hyperboloid is given by
\begin{eqnarray}
V(\tau) =  \frac{\omega^2 R^2}{2}
\, \frac{s_1^2 + s_2^2 +..... + s_{n}^2}{s_0^2}
= \frac{\omega^2 R^2}{2} \coth^2\tau,
\end{eqnarray}
so  for equation (\ref{SCHROD42}) we have
\begin{eqnarray}
\label{OSCIL12}
\frac{d^2 Z}{d\tau^2} + \left[{\epsilon} +
\frac{(L+\frac{n-2}{2})^2-\frac14}{\cosh^2{\tau}}
-
\frac{\nu^2-\frac14}{\sinh^2\tau}\right]\, Z = 0
\end{eqnarray}
where $\nu = \sqrt{\omega^2 R^4+\frac14}$, $\epsilon =
{\tilde E}- \omega^2 R^4$. As in the previous case
the oscillator system is described by the modified P\"oschl-Teller
equation and possesses discrete and continuous spectrum. However,
differing
from the motion on the two-sheeted hyperboloid, the number of bound
states
depends on the total angular momentum.
The discrete state wave functions regular on the line
$\tau \in (-\infty, \infty)$ are
\begin{eqnarray}
\label{OSCIL13}
Z(\tau)
&\equiv& Z_{n_r L}(\tau)
=
\sqrt{\frac{(L-\nu-2n_r+\frac{n}{2}-2)\Gamma(L-n_r+\frac{n}{2}-1)
\Gamma(n_r+\nu+1)}
{R^n (n_r)![\Gamma(\nu+1)]^2\Gamma(L-\nu-n_r+\frac{n}{2}-1)}}
\nonumber\\[2mm]
&\times&
\,
(\sinh\tau)^{\nu+\frac12}
\,
(\cosh\tau)^{2n_r-L-\frac{n}{2}+\frac32}
{_2F_1}(-n_r, - n_r+L+\frac{n}{2}-1; \, \nu+1; \, \tanh^2\tau),
\end{eqnarray}
and
\begin{eqnarray}
\label{OSCIL14}
{\epsilon} = - (2n_r - L + \nu - \frac{n}{2}+2)^2
\end{eqnarray}
where the bound states occur for
$n_r=0,1,.., n_r^{max} = [\frac12(L-\nu+\frac{n}{2}-2)]$.
The last formula means that the discrete spectrum depends on
quantum number $L$ and the energy spectrum of the oscillator
system takes the form
\begin{eqnarray}
\label{OSCIL15}
E_{n_r L} (R) = - \frac{1}{2R^2}
\Bigl[(2n_r-L+2)(2n_r-L-n+3) +
(2\nu-1)(2n_r-L-\frac{n}{2}+2)\Bigr].
\end{eqnarray}

\noindent
{\bf 4.2}
The Coulomb potential on the n-dimensional hyperboloid has
the form \cite{IS,GRO1}
\begin{eqnarray}
V(\tau) = - \frac{\alpha}{R}
\left(\frac{s_0}{\sqrt{s_1^2 + s_2^2 +.....+ s_{n}^2}} +1 \right)
= - \frac{\alpha}{R}\, (\tanh\tau+1).
\end{eqnarray}
The Schr\"odinger equation for this potential is
\begin{eqnarray}
\frac{d^2 Z}{d\tau^2} + \left[({\tilde E}+2\alpha R) +
\frac{(2L+n-1)(2L+n-3)}{4\cosh^2{\tau}} + 2\alpha R \tanh\tau
\right]\, Z = 0,
\end{eqnarray}
which coincides with the {\it Rosen-Morse} equation \cite{LANGE}.

Making the transformation from variable $\tau$
($-\infty<\tau<\infty$), to the new variable $\mu \in [0, \infty)$
\begin{eqnarray}
e^{\tau} = \sinh\mu,
\end{eqnarray}
we go to the equation
\begin{eqnarray}
\label{COOS23}
\frac{d^2 W}{d\mu^2} + \left[({\tilde E} + 4\alpha R) +
\frac{(2L+n-2)^2-\frac14}{\cosh^2\mu} -
\frac{(-{\tilde E}) - \frac{1}{4}}{\sinh^2\mu}
\right]\, W = 0
\end{eqnarray}
where $W(\mu) = (\tanh\mu)^{\frac12} Z(\mu)$.
>From this equation we see that, up to the substution
\begin{eqnarray}
\label{COOS24}
{\tilde E} \to {\tilde E} + 4\alpha R,
\qquad
\nu^2 = - {\tilde E},
\end{eqnarray}
and the simultaneous transformation for total angular momentum
$L \rightarrow 2L$, the quasiradial equation (\ref{COOS23})
for the Coulomb problem on the $n^{coul}=d+1$-dimensional
one-sheeted hyperboloid coincides with the $n^{osc} = 2d$-dimensional
quasiradial oscillator equation (\ref{OSCIL12}).

Comparing now eq. (\ref{COOS23}) with (\ref{OSCIL12}) and taking
into account the eqs.  (\ref{OSCIL14}) and (\ref{COOS24}), we see
that the discrete wave function satisfying the normalization condition
\begin{eqnarray}
R^n\,
\int_{-\infty}^{\infty}
|Z_{n_r L}^n(\tau)|^2 \, d\tau =
R^n\,
\int_{-\infty}^{\infty}
|Z_{n_r L}^n(\mu)|^2 \, \coth\mu \, d\mu = 1.
\end{eqnarray}
has the form
\begin{eqnarray}
\label{OSCIL33}
Z_{n_r L}^n(\tau)
&=&
\frac{2^{n_r-L-\frac{n}{2}}}
{\Gamma(L-n_r+\frac{n-1}{2})}
\,
\sqrt{\frac{[(L-n_r+\frac{n-3}{2})^2-\sigma^2]
\Gamma(2L-n_r+n-2)\Gamma(L+\frac{n-1}{2})}
{R^n (L-n_r+\frac{n-3}{2})(n_r)!
\Gamma(L-\sigma+\frac{1}{2})}}
\nonumber\\[2mm]
&\times&
(\cosh\tau)^{n_r-L-\frac{n-1}{2}}
\,
(e)^{\tau(\sigma-1)}
\nonumber\\[2mm]
&\times&
{_2F_1}\left(-n_r, - n_r+L+ n-2; \, L-n_r+\frac{n-3}{2}+\sigma; \,
\frac{1}{1+e^{-2\tau}}\right),
\end{eqnarray}
with  the discrete energy spectrum of the Coulomb problem
described by the formula
\begin{eqnarray}
E_n = - \frac{(L-n_r-1)(L-n_r+n-2)}{2R^2} -
\frac{\alpha^2}{2(L-n_r+\frac{n-3}{2})^2}-
\frac{\alpha}{R}.
\end{eqnarray}
Bound states occur for
$n_r=0,1,.., n_r^{max} = [(L+\frac{n-3}{2}+\sigma)]$.

Finally note that in distinction to the  sphere and two-sheeted
hyperboloid,
the contraction limit $R\to\infty$ on one-sheeted hyperboloids for
the oscillator and Coulomb problems  makes no sense.

\section*{Acknowledgments}

We thank Professors A.N.Sissakian and  V.M.Ter-Antonyan and Dr.
A.Nersessian for interesting discussions.

\end{document}